\PassOptionsToPackage{table,dvipsnames}{xcolor}

\documentclass[9pt,conference]{IEEEtran}

\usepackage[preprint]{waspaa25} 

\usepackage{bm} 
\usepackage{tikz}
\usetikzlibrary{shapes,arrows.meta, positioning, calc}
\usepackage{soul}

\definecolor{lightlightgray}{gray}{0.93} 

\title{Sound event detection with audio-text models \\and heterogeneous temporal annotations}

\name{Manu Harju,
      Annamaria Mesaros\thanks{This work was supported by Academy of Finland grant 332063 ``Teaching machines to listen". The authors wish to thank CSC-IT Centre of Science Ltd., Finland,  for providing computational resources.}}
\address{Signal Processing Research Centre, Tampere University, Finland \;
}

\begin{document}

\maketitle

\begin{abstract}

Recent advances in generating synthetic captions based on audio and related metadata allow using the information contained in natural language as input for other audio tasks. In this paper, we propose a novel method to guide a sound event detection system with free-form text. We use machine-generated captions as complementary information to the strong labels for training, and evaluate the systems using different types of textual inputs. In addition, we study a scenario where only part of the training data has strong labels, and the rest of it only has temporally weak labels. Our findings show that synthetic captions improve the performance in both cases compared to the CRNN architecture typically used for sound event detection. On a dataset of 50 highly unbalanced classes, the PSDS-1 score increases from 0.223 to 0.277 when trained with strong labels, and from 0.166 to 0.218 when half of the training data has only weak labels. 

\end{abstract}

\section{Introduction}
\label{sec:intro}
Deep learning has recently become the standard approach in various problems in machine listening. One of them is sound event detection (SED), the task of recognizing sound events and when they occur \cite{mesaros:sed}. In unconstrained audio, different class events can overlap, and the same event class can occur multiple times within the same clip, making sound event detection more complicated than audio tagging (AT), where it is sufficient to recognize if a class is present or not. A related task is audio captioning, in which the system outputs a natural language description of the acoustic content of the input.

AT and SED systems are normally trained using supervised learning, which requires annotated data. For audio tagging, it is enough to have temporally weak labels, where the annotation contains just the class labels for each clip. SED requires also temporal information for each event instance, which means that the annotation is more time-consuming; hence, the available datasets with temporally strong labels are smaller than the ones with temporally weak labels.
The largest available dataset containing weakly-labeled audio is AudioSet \cite{gemmeke:audioset}, which consists of approximately 2M 10-second clips from YouTube. A subset of AudioSet was reannotated to include temporal information for sound events \cite{hershey:strong}, but the size of the strongly annotated subset is smaller, around 100k clips. 
Datasets with human-generated audio captions are even smaller, but there have been several studies to produce natural language descriptions with the aid of other information available for the audio clips. For example WavCaps \cite{mei:wavcaps} introduced a method to use ChatGPT to generate and filter audio captions; AudioSetCaps \cite{bai:audiosetcaps} and Sound-VECaps \cite{yuan:soundve} further developed the idea by studying methods to generate captions for AudioSet. 

Training SED systems with a limited amount of strongly annotated data has been studied extensively in the SED task of the past editions of DCASE Challenge \cite{mesaros:decade, turpault:sed, cornell:task4}. In addition to data with temporally strong labels, the task setup included data with temporally weak labels, as well as data without any labels \cite{turpault:sed}. One proposed solution to overcome the lack of sufficient strong labels is the tag-conditioned system by Ebbers and Haeb-Umbach \cite{ebbers:fbcrnn}. The system is a two-stage model, where the first one predicts the tags for the input clip, and the second stage produces the strong labels. 
Another way to solve the problem is to reconstruct the missing strong labels from weak annotations using methods such as active learning \cite{martinsson:weaktostrong}, where a model is used iteratively to refine the labels while updating the model between iterations. Finally, the DCASE Challenge baseline uses a teacher model for self-annotating the missing data during  training \cite{turpault:heterogeneous}.

Audio captions are textual descriptions that contain information about the acoustic content of the audio clips, usually describing the sound events and their attributes or relationships.
The advances in automatic audio captioning provide tools for producing artificial audio captions for any audio. The most recent methods first extract different types of feature tokens, and then fuse the features and generate a caption using a large language model. Finally, the matching of the generated caption and the original audio can be checked with a model such as CLAP \cite{wu:clap} to filter out irrelevant captions. For example AudioSetCaps \cite{bai:audiosetcaps} uses a set of different audio-related features, while the Sound-VECaps pipeline also includes information from video frames \cite{yuan:soundve}. Such methods allow generating rich descriptions for the acoustic content; we hypothesize that we can use this content expressed in natural language to support the sound event detection system. 

Audio captioning and SED have quite different solutions architecturally, though they both aim to output information about the acoustic content of the clip.
Audio captioning systems are usually encoder-decoder networks consisting of an audio encoder and a text decoder. 
Although the audio encoder itself can be derived from an SED system, a more common choice is to use a model pretrained for audio classification (or tagging).
However, an SED system can be used to guide an audio captioning system \cite{xie:sedcaptions}.
Moreover, unsupervised correspondence learning is a sufficiently good training to produce a system capable of phrase-based SED \cite{xie:unsupervised}. 
Finally, in language-queried source separation, text inputs are used to describe the desired separation outputs \cite{liu:separate}.

In this paper, we propose a method to include a text input branch in a CRNN for SED to guide the system in producing the outputs. Our hypothesis is that the auxiliary information contained in the synthetically generated captions helps the model in the absence of strongly-labeled data. Audio-text models for generating captions might lack the ability to place the events in correct temporal order \cite{wu:audiotext}, 
but they can provide valuable additional knowledge in training, and strong prior knowledge during testing. 

The contributions of this paper are: 1) a novel system that includes natural language input in a SED system, and 2) a systematic evaluation of the proposed system using different proportions of strongly labeled, weakly labeled, and automatically captioned data for training.

Our findings show that the CRNN is still a capable architecture for sound event detection, and performs well when all the training data is strongly labeled. Using synthetic captions in training improves the performance of the system with all the text types used in the testing. The effect is similar when only half of the training data is strongly labeled and the other half is only weakly labeled. Furthermore, using synthetic captions in the inference can compensate the drop in performance caused by the missing strong labels in training.



\begin{figure*}
    \centering
    \begin{tikzpicture}[
    audioembedding/.style={rectangle, draw=red!60, fill=red!10, very thick, minimum size=5mm},
    tokenembedding/.style={rectangle, draw=blue!60, fill=blue!10, very thick, minimum size=5mm},
    token/.style={rectangle, draw=purple!60, fill=purple!20, very thick, minimum size=5mm},
    model/.style={rectangle, draw=gray!60, fill=gray!10, very thick, minimum width=20mm, minimum height=5mm},
    sum/.style={circle, thick, draw=black, minimum size=3mm},
    arr/.tip={Latex[angle=30:2.0mm]}
    ]
    
        \node[model] (cnn) {CNN};
        \node (audio) [left of=cnn, node distance=18mm] {$x_a$};
        
        \node[model] (textenc) [below of=cnn, node distance=15mm, align=center] {Text\\encoder};
        \node (text) [left of=textenc, node distance=18mm] {$x_t$};

        \node[sum] (concat) [right of=cnn, node distance=50mm] {$\circ$};
        \node[model] (ca) [right of=textenc, node distance=35mm, align=center] {Cross\\attention};
        
        \node[model] (rnn) [right of=concat, node distance=20mm] {RNN};
        \node[model] (head) [right of=rnn, node distance=25mm] {Linear head};
        \node[model] (pool) [below of=head, node distance=15mm] {Pooling};

        \node (strongpred) [right of=head, node distance=18mm] {$y_s$};
        \node (weakpred) [right of=pool, node distance=18mm] {$y_w$};
        
       
        \draw[-arr] (audio) -- (cnn);
        \draw[-arr] (text) -- (textenc);

        \draw[-arr] (cnn) -- (concat);
        \draw[-arr] (cnn) -| node[pos=0.9,anchor=east]{Q} (ca);
        \draw[-arr] (textenc) -- node[pos=0.6,anchor=north]{K, V} (ca);
        \draw[-arr] (ca) -| (concat);
        
        \draw[-arr] (concat) -- (rnn);
        \draw[-arr] (rnn) -- (head);

        \draw[-arr] (head) -- (pool);

        \draw[-arr] (head) -- (strongpred);
        \draw[-arr] (pool) -- (weakpred);
    
    \end{tikzpicture}
    \vspace{-15pt}
    \caption{Block diagram of the proposed system. In case where text inputs $x_t$ are not provided, the text encoder and cross attention are ignored and the audio features corresponding to the audio input $x_a$ are concatenated with zeros.}
    \label{fig:modelblocks}
\end{figure*}
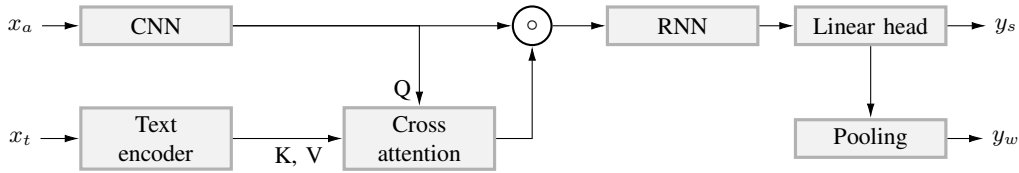


\section{Proposed method}
\label{sec:method}


Our proposed model loosely follows the baseline architecture introduced in the DCASE Challenge 2019 Sound event detection task \cite{turpault:sed}. The block diagram for the proposed system is shown in Figure \ref{fig:modelblocks}. Our implementation is also available on GitHub\footnote{https://github.com/mnuhurr/text-sed}.

The upper data path in the figure follows the traditional CRNN architecture: audio features $x_a$ are fed into a 2D CNN, followed by an RNN, and finally into a linear head to produce frame-wise predictions $y_s$. We use a CNN consisting of 6 convolutional blocks using gated activations. For the RNN we use two-layer bidirectional GRU. The output of the recurrent part is projected to frame-wise class predictions using a single linear layer. 

To handle additional text inputs $x_t$, we add a transformer-based text encoder and a single cross-attention for fusing the features. The CNN outputs are used as the query, and the text encoder outputs are used as the keys and values. The cross-attention output is concatenated with the CNN output along the feature dimension before being fed into the RNN part. In cases where text inputs are not provided, the text encoder and cross-attention are ignored, and the audio features are concatenated with zeros.
Another way to connect the text branch is to add the cross-attention outputs to the audio features produced by the CNN. However, based on preliminary experiments, we opted to use concatenation.
In several experiments we use weakly labeled training data. For the sake of simplicity, the model produces weak labels by pooling the frame-wise outputs over time. In this work we use the arithmetic means of average and maximum of the temporally strong logits for each class.

The network takes log-mel spectrograms for the audio input, for which we use 64 mel bands. The audio is resampled to 16 kHz sampling rate; we use 25 ms windows with 20 ms hop size. 
To overcome the high imbalance of the classes, we use data resampling in the training. We also use SpecAugment \cite{park:specaugment} and spectrogram mixing \cite{niizumi:byol} for data augmentation, as it was generally shown that data augmentation improves generalization of SED models.

The text inputs are processed using a BPE tokenizer \cite{sennrich:neural} with a vocabulary size of 6000 tokens, and the vocabulary is constructed using the available training captions. The text encoder is a standard transformer \cite{vaswani:attention} with 4 layers and 8 heads. While it is possible to use a pretrained model for the text encoder, for this study we train it from scratch. 
This gives a better picture of the learning and of the importance of the text branch in the proposed system.
By training it for this task specifically, we can keep the text branch lightweight and the number of weights comparable to the other parts of the model.

\section{Experiments}
\label{sec:experiment}

\begin{table*}
\begin{center}
    \begin{tabular}{cc|l|cc|ccc}
        \multicolumn{2}{c}{Training data} \\
        Captions & Strong / Weak & Eval text & PSDS-1* & PSDS-2* & F-score & Precision & Recall \\
        \midrule
        \midrule
        \rowcolor{lightlightgray}
        - & 100 / 0 & -            & 0.223 & 0.341 & 16.5 & 38.3 & 10.5 \\
        - & 100 / 0 & AudioCaps    & 0.219 & 0.336 & 16.3 & 37.6 & 10.4 \\
        
        \midrule
        \checkmark & 100 / 0 & -               & 0.180 & 0.280 & 14.5 & 39.8 &  8.9 \\
        \checkmark & 100 / 0 & Tags            & 0.252 & 0.396 & 16.4 & 44.0 & 10.1 \\
        \checkmark & 100 / 0 & AudioCaps       & 0.246 & 0.378 & 16.6 & 42.8 & 10.3 \\
        \checkmark & 100 / 0 & Sound-VECaps    & 0.277 & 0.417 & 18.0 & 44.1 & 11.3\\

        \midrule
        \midrule
        \rowcolor{lightlightgray}
        - & 50 / 50 & -            & 0.166 & 0.281 & 16.2 & 24.4 & 12.1 \\
        - & 50 / 50 & AudioCaps    & 0.161 & 0.273 & 16.1 & 23.3 & 12.3 \\

        \midrule
        \checkmark & 50 / 50 & -               & 0.120 & 0.212 & 14.4 & 22.5 & 10.6 \\
        \checkmark & 50 / 50 & Tags            & 0.181 & 0.309 & 15.6 & 25.4 & 11.3 \\
        \checkmark & 50 / 50 & AudioCaps       & 0.183 & 0.313 & 16.3 & 26.2 & 11.9 \\
        \checkmark & 50 / 50 & Sound-VECaps    & 0.218 & 0.350 & 17.9 & 28.0 & 13.1 \\
    \end{tabular}
    
    \vspace{6pt} 
    \caption{CRNN results when evaluated using corresponding audio-text pairs. Upper half of the table shows results in the setting where all the training data has strong labels, and the lower half contains results for the case where half of the training data has only weak labels. The rows with highlighted gray background are the base cases without any text in training and testing. 
    }
    \label{tab:numbers}
\end{center}
\end{table*}

\begin{figure}[!t]
    \centering
    \includegraphics[width=0.95\linewidth]{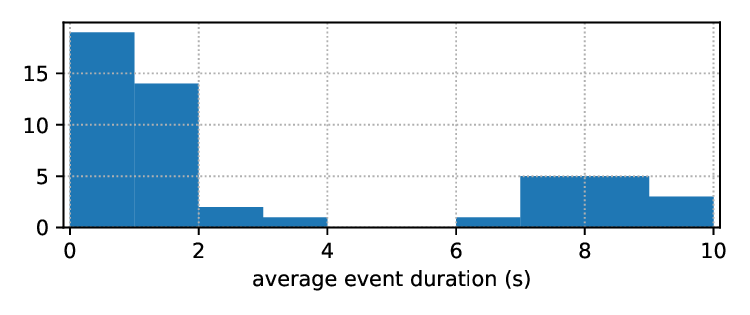}
    \vspace{-10pt}
    \caption{Average event duration of different classes in the training data. 33 classes have average event duration shorter than 2 seconds, while the longer ones may be active over the entire 10 seconds and continue beyond the clip boundaries.
    }
    \label{fig:avgdur}
\end{figure}

\subsection{SED Data}
We use the strongly annotated subset of AudioSet \cite{hershey:strong} as our audio data and event annotations, and Sound-VECaps \cite{yuan:soundve} as training captions. For evaluation we also want to include human-generated captions. AudioCaps \cite{kim:audiocaps} is a subset of the weakly labeled AudioSet, consisting of a training set of 49838 files with one caption each and smaller validation and test sets with five captions for each file. The intersection of the training subsets of AudioCaps and AudioSet-strong contains 9028 files, which forms the basis of our test set.
In other words, for training and validation, we use those files in the training split of AudioSet-strong which do not have a caption in the training split of AudioCaps.
For simplicity, we only consider the training split of AudioCaps, which leaves out 89 files from the validation set and 178 files from the test set.

We choose 50 most frequent tags in our training split for the event classes; they include  specific classes such as ``Whistling'' and ``Baby cry, infant cry'', as well as more general and ambiguous ones, such as ``Human sounds'', ``Mechanisms,'' and ``Generic impact sounds''. Moreover, there are classes in hierarchical relation, e.g. ``Dog'' and ``Bark''. The classes are highly imbalanced: the total duration of ``Music'' is more than 46 hours in the training split, while the smallest class ``Dishes, pots and pans'' has less than 8 minutes of annotated events. 
There is also a large imbalance in the distribution of the event durations: some classes like ``Music'' and ``Train'' tend to last over the whole clip if they are present, whereas most of the other included classes are annotated in very short events. 
The histogram of the average event durations of different classes in the training data is shown in Figure \ref{fig:avgdur}: 33 of the 50 classes have average event duration under 2 seconds, and there are 14 classes with longer average durations. Furthermore, events annotated to start or end at the clip borders are likely to continue outside the annotated clip, thus the event durations are truncated.



Due to the data originating from YouTube, some clips originally included in AudioSet-strong are not anymore available. After filtering out the files with no sound events in the selected 50 largest classes, our final dataset sizes are 69744, 7750, and 8301 files for training, validation, and testing, respectively.

\subsection{Synthetic and Manually Annotated Captions}
For the synthetic captions we use the filtered version of Sound-VECaps \cite{yuan:soundve}, which are reported to contain only audible content.
The synthetic captions in the training data are long compared to human-annotated captions: the average caption length in AudioCaps is 8.7 words, whereas the average caption length in Sound-VECaps is 31.3. These captions were generated using also information extracted from video frames, and despite filtering, the audio-only version of the data still has some captions containing expressions that can only be generated by visually inspecting the scene. For example, ``Music is playing while a camera captures the urban scene, with a ferris wheel, a boat, a hospital building, and a park in the background, --'' cannot be generated using only acoustic information.

\subsection{System training and evaluation}

The models are trained in a supervised setting using log-mel spectrograms and possibly text tokens as inputs. In the first experiment, all the training data is strongly labeled; in the second part, half of the training data has only weak labels. In the scenarios where text is used in the training, the text inputs are dropped out with 50 \% probability to prevent the model from relying too much on the text branch.

When evaluating the models with both input modalities present, we use three kinds of text data: human-generated captions (AudioCaps), lists of tags (AudioSet-strong as tags, class labels taken from the AudioSet ontology), and synthetic captions (Sound-VECaps). For each scenario, we train and evaluate five models, and the presented numbers are averages over the five runs.

To investigate the robustness of the proposed system to incorrect texts, we also evaluate the models using tags where we include randomly chosen incorrect tags. For each number of incorrect tags, each model is evaluated five times. The incorrect tags are randomized each time, as well as the order in which the tags are provided to the system. We also evaluate the models by leaving out one tag at a time.

We measure performance using two different methods: polyphonic sound detection score (PSDS) \cite{bilen:psds} and segment-based F-score \cite{mesaros:metrics}. As in DCASE Challenge 2021-2023 Sound Event Detection tasks, we use PSDS-1 and PSDS-2 to measure two aspects of the detection performance: PSDS-1 is more sensitive to temporal accuracy, whereas PSDS-2 measures more class confusions. 
We change the parameter $\alpha_{ST}$ from the DCASE setup where $\alpha_{ST}=1$ and penalizes the inconsistencies in the true positive rates across the classes. The 50 classes in our experiments form a diverse and highly unbalanced set across the AudioSet class hierarchy, hence we expect the models to perform differently on the different classes. For this reason we set $\alpha_{ST} = 0$ for both PSDS metrics to remove this cost of instability across classes. A similar approach was taken e.g. in \cite{schmid:effective} for the same reason. 
We use asterisk in the metric names to denote this difference to the DCASE Challenge PSDS scenarios. 

The segment-based F-score is computed using one-second segments. For the F-score we need to binarize the system outputs, and we simply use a 0.5 threshold in all the scenarios.

\section{Results and discussion}
\label{sec:resdisc}

\begin{figure*}
\centering
    \vspace{-10pt}
    \includegraphics[width=0.95\linewidth]{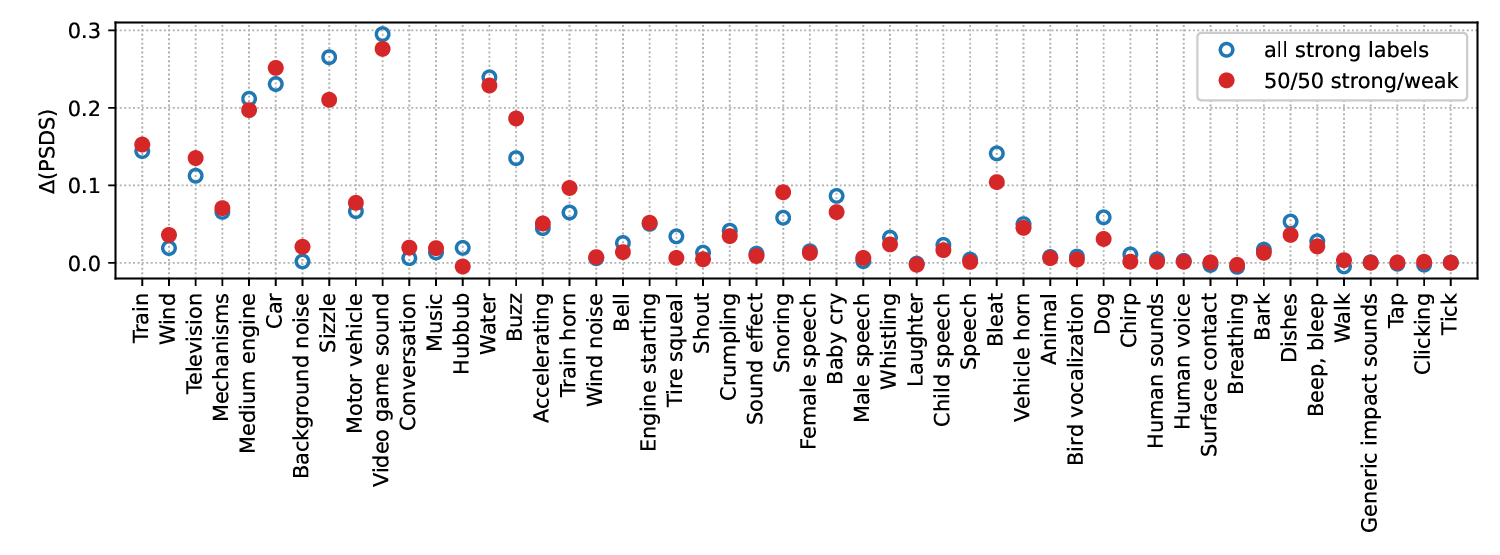}
    \vspace{-10pt}
    \caption{Class-wise change in PSDS-1 between the baseline CRNN without text and the system using synthetic captions in training and evaluation. Blue markers are for when all the training data has strong labels, and red markers when half of the training data is only weakly labeled. The classes are ordered by the average event duration. Some of the class names are shortened from the original AudioSet labels to make the figure smaller.}
    \label{fig:effect}
\end{figure*}


The evaluation results with different types of corresponding textual inputs are shown in Table \ref{tab:numbers}. The table also includes the baseline case where the model is trained and tested with no text data. The base case rows are highlighted with gray background. 



The first row in Table \ref{tab:numbers} corresponds to the baseline CRNN-based SED, a basic setup without any text in training or evaluation; the evaluation using AudioCaps on the following row is included to see how much an untrained text encoder affects the outputs. 
The baseline CRNN achieves PSDS-1* 0.223 when evaluated in the simple SED setting without text inputs. The proposed method scores PSDS-1* 0.252 when evaluated using lists of tags, and 0.277 with synthetic captions. The increase in the scores with tags and human-generated captions is over 10\% (0.029 and 0.023, respectively), and with the synthetic captions over 20\% (0.054). For PSDS-2* the effect is similar: baseline CRNN achieves 0.341, and the model trained with text scores higher when evaluated with all the used text types: 0.396, 0.378, and 0.417 with tags, human-generated captions, and the synthetic captions, respectively. 

The effect on the segment-based F-score is not as visible: the F-score is around the baseline level with tags and human-generated captions, but increases slightly with the synthetic captions. However, precision improves significantly when text inputs are used. 

The second row in the table contains the numbers for the model trained without text, but evaluated with human generated captions. In the optimal case, the model should learn to ignore the text branch, and the numbers should be the same as without the texts. There are small differences in the score, but in practice the model learns to produce the outputs only using the audio features.

In the case when half of the training data lacks strong labels (shown as 50/50 rows in Table \ref{tab:numbers}), the PSDS-1* and PSDS-2* scores of the baseline CRNN drop to 16.6 and 28.1, respectively. Using the text inputs in training and evaluation results in similar improvements as in the first experiment. The results show that with the synthetic captions, the proposed method is able to compensate for the performance drop caused by the missing strong annotations.

\subsection{Detailed performance analysis}

Figure \ref{fig:effect} shows the class-wise differences in PSDS-1 between the proposed method and the baseline CRNN. The classes are ordered by the average event duration. We can clearly see that for most of the classes there is an improvement in performance, with the largest improvements seen for the classes with longer events.

\begin{figure}
    \centering
    \includegraphics[width=1.0\linewidth]{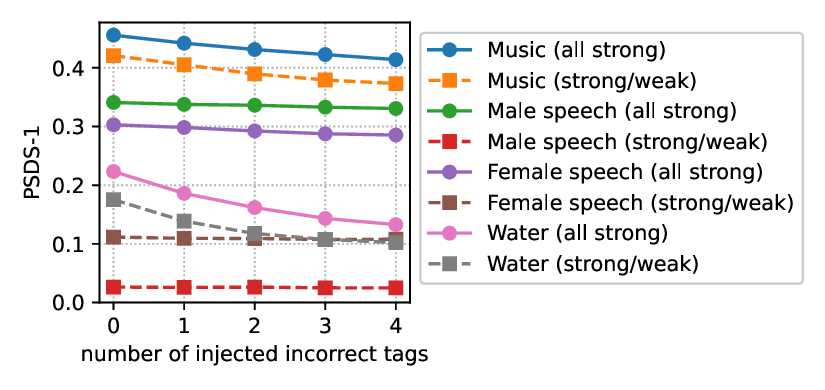}
    \vspace{-20pt}
    \caption{The effect of injecting incorrect class labels for selected classes. The text input includes the list of correct class labels, and a number of random incorrect labels.}
    \label{fig:errtags}
\end{figure}

Table \ref{tab:injectedtags} shows the drop in overall PSDS-1*, when the models are evaluated using lists of tags that include a number of incorrect ones. Each trained model is evaluated five times, and for each run the incorrect tags are randomized; scores are average over all the runs. The drop in PSDS-1* between the zero incorrect tags and five incorrect tags is 15\% (0.038) when all the training data is strongly labeled, but 20\% (0.038) when half of the data is weakly labeled. This hints that the model learns to rely more on the text branch when part of the training data lacks strong labels.

Figure \ref{fig:errtags} shows the PSDS-1 drop for selected classes, when the model is evaluated using tags as the input text, and the tag list also includes randomly selected incorrect tags. The figure shows that moving to partly weakly labeled training data has a modest effect on classes like ``Music'' or ``Water'', but the performance on both speech classes drops quite drastically. On the other hand, adding more incorrect tags has a smaller effect on the speech classes than on ``Music'' and ``Water''. The 1-second segment-based F-scores for the speech classes even increased slightly in the base case when half of the training data is weakly labeled, but this is mostly due to a change in precision/recall balance. 

Leaving out one tag from the input list has the effect of decreasing recall by an average of 8.7 when all the training data is strongly labeled, and by 9.0 when half of the training data has weak labels. Precision increases by 1.2 and 0.8, respectively. This indicates that, as intended, the system learns to use the information in the input lists, as the performance generally drops when a tag is omitted.

\begin{table}[t]
    \centering
    \begin{tabular}{c|cccccc}
        & \multicolumn{6}{c}{Number of incorrect tags} \\
        Strong / Weak & 0 & 1 & 2 & 3 & 4 & 5 \\
        \midrule
        100 / 0   & 0.252 & 0.243 & 0.234 & 0.225 & 0.219 & 0.214 \\
        50 / 50   & 0.181 & 0.171 & 0.161 & 0.154 & 0.148 & 0.143 \\
    \end{tabular}
    \vspace{5pt}
    \caption{Overall PSDS-1* values for the systems evaluated with lists of correct tags and $N$ randomly chosen incorrect tags.}
    \label{tab:injectedtags}
\end{table}

\subsection{Discussion}


Different types of audio-text models exist in literature but, to our best knowledge, this is the first work on using text to assist sound event detection. 
The results show that even by training from scratch, it is possible to learn a language model that benefits the SED-performing CRNN. Moreover, at inference, the model can take advantage of different types of text inputs, not necessarily the kind used in training, varying from just simple label names to long synthetic descriptions. 

Using the cost of instability across classes $\alpha_{ST} = 1$, the baseline CRNN and the proposed method only reach PSDS-1 scores 0.048 and 0.054, respectively. Compared to the numbers in Table \ref{tab:numbers}, the penalization caused by $\alpha_{ST}$ is remarkable. 
When half of the training data lacks strong labels, but the system is trained and evaluated using the synthetic captions, the PSDS-1* and PSDS-2* scores are comparable to the baseline CRNN scores using strong labels. However, the penalized PSDS scores do not gain such increase, as the system performance improves mostly on already better scoring classes. 

\section{Conclusions}
\label{sec:concl}

This study presented a framework to include a text guiding branch to a sound event detection system, in an effort to include additional information in training and priors to the inference. The proposed method uses machine-generated captions for training, hence it does not require any additional annotation. Moreover, the  proposed method can leverage weakly-labeled data as additional source of information in training, complementing the strongly-labeled data and synthetic captions. In practice, the method increases the amount of training data for the system, offering a simple and effective solution to the scarcity of strongly-labeled data available for training SED systems. 



\clearpage
\IEEEtriggeratref{12}
\bibliographystyle{IEEEtran}
\bibliography{waspaa_paper}

\end{document}